\documentclass[sigconf]{acmart}

\usepackage{graphicx}
\usepackage{framed}

\usepackage{footmisc}
\usepackage{dirtytalk}

\AtBeginDocument{%
  \providecommand\BibTeX{{%
    \normalfont B\kern-0.5em{\scshape i\kern-0.25em b}\kern-0.8em\TeX}}}

\copyrightyear{2022} 
\acmYear{2022} 
\setcopyright{acmlicensed}\acmConference[EASE 2022]{The International Conference on Evaluation and Assessment in Software Engineering 2022}{June 13--15, 2022}{Gothenburg, Sweden}
\acmBooktitle{The International Conference on Evaluation and Assessment in Software Engineering 2022 (EASE 2022), June 13--15, 2022, Gothenburg, Sweden}
\acmPrice{15.00}
\acmDOI{10.1145/3530019.3530024}
\acmISBN{978-1-4503-9613-4/22/06}

\begin{document}

\title{How Are Communication Channels on GitHub Presented to Their Intended Audience? -- A Thematic Analysis}
\author{Verena Ebert}
\email{verena.ebert@iste.uni-stuttgart.de}
\affiliation{%
  \institution{University of Stuttgart}
  \city{Stuttgart}
  \country{Germany}
 }
 
\author{Daniel Graziotin}
\email{daniel.graziotin@iste.uni-stuttgart.de}
\affiliation{%
  \institution{University of Stuttgart}
  \city{Stuttgart}
  \country{Germany}
 }
 
\author{Stefan Wagner}
\email{stefan.wagner@iste.uni-stuttgart.de}
\affiliation{%
  \institution{University of Stuttgart}
  \city{Stuttgart}
  \country{Germany}
}


\begin{abstract}
Communication is essential in software development, and even more in distributed settings. Communication activities need to be organized and coordinated to defend against the threat of productivity losses, increases in cognitive load, and stress among team members. With a plethora of communication channels that were identified by previous research in open-source projects, there is a need to explore organizational issues in how these communication channels are introduced, explained, and motivated for use among all project members.
In this study, we wanted to understand which communication channels are used in GitHub projects and how they are presented to the GitHub project audience. 
We employed thematic analysis to analyze 151 artifacts in 90 GitHub projects.
Our results revealed 32 unique communications channels that can be divided into nine different types. Projects mostly provide channels of different types, but for some types (e.g.,\ chat) it is common to provide several channels. Maintainers are aware that channels have different properties and help the developers to decide which channel should be used in which case. However, this is not true for all projects, and often we have not found any explicit reasons why maintainers chose to provide one channel over another. 
Different channels can be used for different purposes and have different affordances, so maintainers have to decide wisely which channels they want to provide and make clear which channel should be used in which case. Otherwise, developers might feel overwhelmed of too many channels and information can get fragmented over multiple channels~\cite{storey}. 
\end{abstract}

\begin{CCSXML}
<ccs2012>
   <concept>
       <concept_id>10003120.10003130.10003233.10003597</concept_id>
       <concept_desc>Human-centered computing~Open source software</concept_desc>
       <concept_significance>500</concept_significance>
       </concept>
   <concept>
       <concept_id>10011007.10011074.10011134</concept_id>
       <concept_desc>Software and its engineering~Collaboration in software development</concept_desc>
       <concept_significance>500</concept_significance>
       </concept>
 </ccs2012>
\end{CCSXML}

\ccsdesc[500]{Human-centered computing~Open source software}
\ccsdesc[500]{Software and its engineering~Collaboration in software development}

\keywords{GitHub projects, communication channels, open-source}

\maketitle

\section{Introduction}

Several studies have shown that software developers spend most of their working days communicating~\cite{perry1994, herbsleb1995, astromskis2017}. A big part of communicating is choosing communication media~\cite{DeSaLeitaoJunior2019}. Communicating to this extent can be even overwhelming for developers~\cite{Whittaker_1996}, leading to important messages getting lost~\cite{Whittaker_1996} or even broken builds~\cite{Damian2007AwarenessOccur}.

Additionally, there are other problems. Communication bits can be scattered over different communication channels~\cite{storey2014, gutwin2004}. With this, software projects face a substantial loss of development speed~\cite{herbsleb1999,herbsleb1995}, to the point that communication failures are seen as the cause of productivity losses~\cite{herbsleb2003} or even project failures~\cite{Begel2014}. Furthermore, using several communication channels can be disruptive~\cite{Storey2016, storey} and developers have to face many obstacles when communicating like facing different cultures, trust issues, different time-zones or language problems~\cite{Olson2000}.

In open-source software (OSS) projects, which are characterized by distributed environments, a good communication system is of particular importance. Team members have to coordinate work, discuss tasks, solve problems, make decisions, and manage their projects continuously, over different time zones~\cite{bird2008}.

In general, using communication channels is nothing new. Developers use many channels. Previously, mailing lists were very common~\cite{shihab2010}. Nowadays, other communication channels become more and more popular~\cite{Kafer2018CommunicationEra, storey}, for example, Slack~\cite{lin2016}, issue trackers~\cite{guzzi2013},  Twitter~\cite{Singer2014} or Gitter~\cite{Parra2022}.

One of the largest platforms for open-source software developments is GitHub. In 2021, there were over 200 million projects on GitHub and over 73 million developers.\footnote{\url{https://github.com/}}
On GitHub, not only the maintainers of a project can contribute code, but also other developers can fork a project and then submit a pull request with their changes. Of course, all these developers must somehow communicate using communication channels. However, the official recommendations by GitHub about communication in a project only focuses on GitHub internal channels such as issues or GitHub Discussions\footnote{\url{https://docs.github.com/en/get-started/quickstart/communicating-on-github}} and does not include any recommendations for other channels.

Storey et al.~\cite{storey} surveyed 1,449 GitHub users to see which communication channels are essential for their work and what challenges they face. They found that on average, developers used 11.7 channels across all activities. The five most important reported channels were code-hosting sites, face-to-face, Q\&A sites, web search and microblogs. Mentioned challenges included getting distracted or interrupted by activities in the communication channels. Another challenge was that developers need to keep up with which technology: they have to know which channel is the most important one at the moment. Developers have to be literate with the used channels and also have to face friction of channels, for example that some channels do not offer mobile support or have annoying notifications. Between channels, there was also the problem of information fragmentation and the overwhelming quantity of shared information.

\subsection{Problem Statement}

There is a lack of knowledge on how communication channels are presented in large GitHub projects. With so many communication channels provided in GitHub projects, there is a need to explore how these communication channels are introduced, explained, and motivated in the project documentations. We need to know which types of communication channels are provided and how maintainers introduce them to the developers. For this, we decided to focus on large GitHub projects as defined in section~\ref{sec:def}. GitHub is not only used for developing software in a team, but also for hosting small projects with only one developer who does not communicate with anybody or even just as backup for any non-software-related data. As we want to examine communication channels, we therefore decided to only include large projects, assuming that they will include more communication channels than one-developer-projects.
  
Our results can then be used as a basis for further analyses exploring how GitHub users feel about the current setup and lead to better guidelines for GitHub maintainers regarding provided communication channels.

\subsection{Research Objective}

We want to use thematic analysis as described by Braun and Clarke~\cite{Braun2006} to explore which channels are provided in large GitHub projects and how they are presented. We want to see how the setup of communication channels looks like. 

Following an inductive approach, we wanted to be as open-minded as possible. From another study~\cite{Kafer2018CommunicationEra}, we already knew that projects provide a large variety of channels. Additionally, we wanted to explore any other issue related to communication channels in GitHub projects. 

 \begin{quote} \textbf{RQ 1: What types of communication channels do GitHub projects provide?} \end{quote}

 \begin{quote} \textbf{RQ 2: How are communication channels on GitHub presented to their intended audience?} \end{quote}

\section{Related Studies}
\label{sec:communication}

Besides Storey et Al., there are several other studies about how developers communicate in GitHub projects. 

K\"{a}fer et al.~\cite{Kafer2018CommunicationEra} analyzed communication channels in GitHub projects. They have shown that only half of the analyzed projects use externally visible communication channels. The most often provided channels were GitHub Issues, personal e-mail, Gitter, Twitter and mailing lists. 

Tantisuwankul et al.~\cite{Tantisuwankul2019} showed that contemporary GitHub projects tend to offer multiple communication channels. These channels are constantly changing, meaning that new channels get adopted, and other channels are not used anymore.  However, they focused on the GitHub-internal communication channels like issues or wiki. 

Constantino et al. have interviewed 12 GitHub developers about collaborative software development~\cite{Constantino2020UnderstandingStudy}. Regarding communication channels, they showed that \say{a more significant number of tools does not equate to better communication}. Different channels can lead to failing to reach each developer or providing unsolicited information to non-interested parties. 

Studies are analyzing communication in development teams who are, however, not using GitHub. We report on three of them. 

De S{\'{a}} Leit{\~{a}}o J{\'{u}}nior et al.~\cite{DeSaLeitaoJunior2019} analyzed communication in distributed software development teams using Grounded Theory. One of the theoretical categories they found is \textit{choosing communication media}, including adopting email, video calls, instant messaging software or issue tracking software. 

Snyder and Lee-Partridge~\cite{Snyder2013} analyzed communication channel choices in teams. They showed that participants tend to use face-to-face communication, telephone, or e-mail. Participants chose the used channel based on the information they wanted to share. Additionally, ease of use, reliability, and convenience played a big role in choosing a channel. 

Josyula et al.~\cite{Josyula2019} have identified nine information needs that developers have, e.g., clarifying requirements or understanding and resolving bugs. These needs can be fulfilled by using one of thirteen information sources. Some of these sources are communication channels including blogs and community forums, discussions with colleagues or social networking sites. 

All three studies show that choosing a communication channel is a big part of communication in development teams in general.  Our study will show if maintainers are aware that they can support developers in choosing the right channel by giving information about which channels are supported and which channel is useful in which situation. 

Furthermore, Storey et al.~\cite{storey} have provided six recommendations, based on the challenges that we mention in the introduction of the present paper, that maintainers should follow in their projects. We aim to compare our results to these recommendations in section~\ref{sec:discussion}.

\section{Methodology} 
\label{sec:thematic_analysis}

For this study, we performed a thematic analysis as described by Braun and Clarke~\cite{Braun2006}. \say{Thematic analysis is a method for identifying, analysing, and reporting patterns (themes) within data. It minimally organises and describes your data set in (rich) detail}~\cite{Braun2006}. It is a form of analysis within qualitative research. Other than Grounded Theory, it requires waiting until all data has been collected before coding begins and does not return to data collection once the coding has begun. As we have a fixed data set, thematic analysis was more suitable for us. 

Thematic analysis can be performed in two ways -- inductive and deductive. A deductive approach follows previous knowledge and answers a \say{quite specific research question}~\cite{Braun2006}. Using an inductive approach, on the other hand, allows for \say{coding the data without trying to fit it into a pre-existing coding frame, or the researcher's analytic preconceptions}~\cite{Braun2006}.
We did not have any deep insights or assumptions about how communication channels are presented in GitHub projects, except for some experiences as developers. Therefore, we decided to follow an inductive approach and tried to be as open as possible for anything communication-channel-related that we found in the GitHub projects. 

We also decided to follow a semantic approach. \say{With a semantic approach, the themes are identified within the explicit or surface meanings of the data and the analyst is not looking for anything beyond what a participant has said or what has been written}~\cite{Braun2006}. We expected the data to be rather unemotional texts, so we saw no need for interpreting any underlying assumptions or ideologies, as would be the case with a latent approach. 

We then followed the six phases as proposed by Braun and Clarke~\cite{Braun2006}, namely \say{familiarising yourself with your data, generating initial codes, searching for themes, reviewing themes, defining and naming themes, and producing the report}.

There is some work about thematic analysis in software engineering by Cruzes and Dyb{\aa}~\cite{Cruzes2011}. However, they focus only on finding themes and patterns in primary studies as a way of systematic literature reviews and not on other types of data~\cite{Cruzes2011}. Therefore, we decided not to follow their proposed steps but to keep to the original six steps of thematic analysis as described by Braun and Clarke~\cite{Braun2006}.

\subsection{Definitions}

During our study and in this paper, the following definitions are used:

\begin{description}
\label{sec:def}

\item[\textbf{Communication Channel:}] \hfill \\
A communication channel is \say{a system or method that is used for communicating with other people}\footnote{\url{https://dictionary.cambridge.org/de/worterbuch/englisch/channel-of-communication}}. In this study, communication channels can be uni- or bidirectional. They are used by maintainers, developers and users to communicate with each other.

\item[\textbf{Maintainer:}] \hfill \\
Maintainers are the GitHub users who have the right to merge pull requests of a GitHub project. This group includes every other GitHub role that has more rights in a project, for example, admin. Maintainers set up projects and provide communication channels for other maintainers, developers and users. 

\item[\textbf{Developer:}] \hfill \\
Developers are GitHub users who actively contribute to a project by creating pull requests. 

\item[\textbf{End user:}] \hfill \\
End users are every other person who somehow use the provided communication channels. End users use the software developed in a GitHub project but do not help develop it by adding code. 

\item[\textbf{Large project:}] \hfill \\
Based on~\cite{destefanis:2018, ding:2018, werder:2018}, large GitHub projects match the following criteria:
\begin{itemize}
    \item > 24 commits
    \item $\geq$ 50 comments in total (issue and commit)
    \item > 5 participants
    \item > 2 commits in average /month
    \item > 5 months activity (commits)
\end{itemize}

We decided to exclude projects which are forks of other projects. In another study, we saw that there are fork projects which match the criteria for large projects, but don't provide any communication channels as the main projects are responsible for this. 

\end{description}

\subsection{Data Collection} 

We used the GitHub Torrent database\footnote{\url{https://ghtorrent.org/}} to find large GitHub projects. We used the latest database dump from June 2019 and queried the database using SQL. From originally 1,486,232 projects we had 85,377 projects left matching our criteria for large projects. 
In thematic analysis, as described by Braun and Clarke, there are no hard criteria of how many artifacts should be coded to reach saturation. Instead, the sample should be large enough to get a \say{clear conceptualisation of what those themes represent, and how and why we treat them as significant} while being not too large to not miss the \say{nuance contained within the data}~\cite{Braun2016}. 
Given that analyzing all 85,377 projects would not be feasible, and given the absence of established criteria for a cut-off value, we opted to \textit{start} with an arbitrary value of 100 randomly selected projects to code. 
If during our analysis we had had the impression that data is missing to distinguish the themes clearly, we could have easily added more projects. However, during our analysis, we came to the conclusion that our sample was big enough to define our themes, as we could clearly distinguish all themes before finishing coding the last projects. 

Of these 100 projects, 21 were no longer available on GitHub or marked as shut down by the maintainers when we performed our coding in 2021, leaving 79 projects for our analysis. As the database dump was from 2019, the projects in our list might have changed during the last two years, potentially not fulfilling some criteria for large projects anymore. We could check on most of the criteria, with the only one left out being the \textit{> 2 commits in average /month}. Given that our main analysis does not follow on the frequency of commits per month, we opted to relax this latter criterion and opted to enrich our dataset differently, as follows.

To enrich our dataset in terms of diversity, we looked for outlier projects, that is, projects with a large amount or only a few communication channels. For this, we used a tool which uses regular expressions to mine for communication channels in the README and wiki files of GitHub projects that was developed for a previous study~\cite{Kafer2018CommunicationEra}. We randomly selected
10,000 projects from our original dataset of large projects and analyzed them with this tool. We 
sorted the sampled 10,000 projects by amount of unique channels. Then, we selected one project with no mentioned communication channels plus the ten projects with the largest amount of unique communication channels. We added these selected projects to our dataset. We did not add any projects with only a few channels, as we observed that they were already represented in our first 79 projects. 

Our final dataset comprised 79 randomly selected projects plus the 11 projects with a large amount or only a few communication channels, with a total of 90 projects. Of each project, we analyzed artifacts consisting of the markdown files, the wiki, and the websites if they were referenced in the GitHub project. The relevant website contents were summarized into one file. All together, we coded 151 artifacts.

\section{Data Coding}
The first author collected the data, performed the initial coding and developed the first themes. The other two co-authors discussed codes and the themes collectively and iteratively until the final codes and themes were defined. The first author drove the coding activities, while the other two co-authors suggested improvements, checked on snippets and codes, and provided directions to observe.

\section{Results}

\begin{figure*}[ht]
\centering
\includegraphics[width=0.75\linewidth]{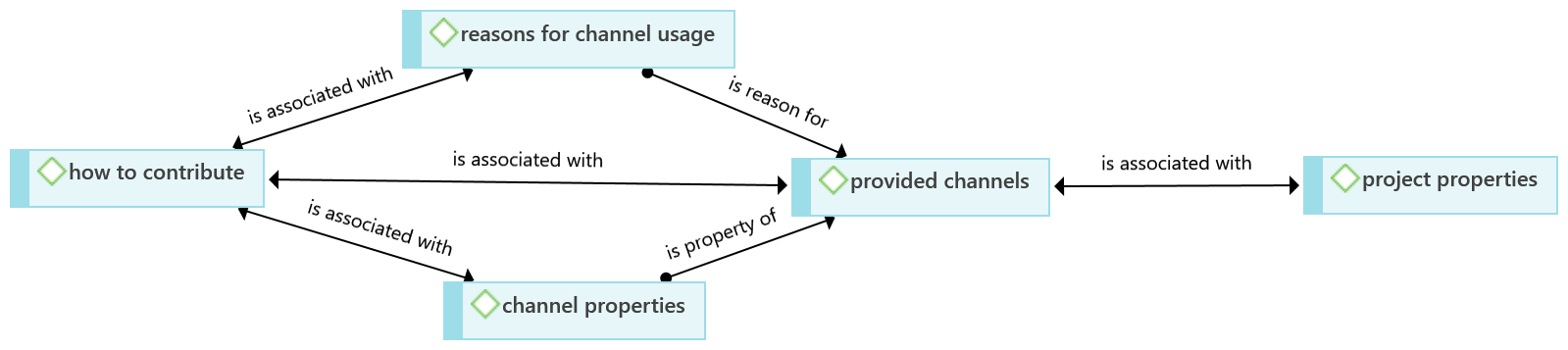}
\caption{Overview of themes}
\label{fig:themes}
\end{figure*}

In the following, we will describe the themes we defined. In the figures, each theme has a blue background, sub-themes have a white background. We will give examples from our analyzed projects for most themes. 
Figure~\ref{fig:themes} shows an overview of the five themes that we defined. Each theme will be described in detail in the following subsections.

\subsection{Themes}

Every GitHub project we analyzed provided some sort of communication channel (\emph{provided channels}). In 22 projects, they used only the by GitHub provided issue tracker and pull requests. In three projects, there was no README file or any other markdown file with additional information for developers or users. We summarized this as (\emph{project properties}). Some project maintainers give reasons why a certain channel should be used (\emph{reasons for channel usage}). Each channel has a set of properties (\emph{channel properties}) and some channels need some explanations or fulfilled preconditions before they can be used (\emph{how to contribute}). 

\subsection{Provided channels}

\begin{figure*}[ht]
\centering
\includegraphics[width=0.85\linewidth]{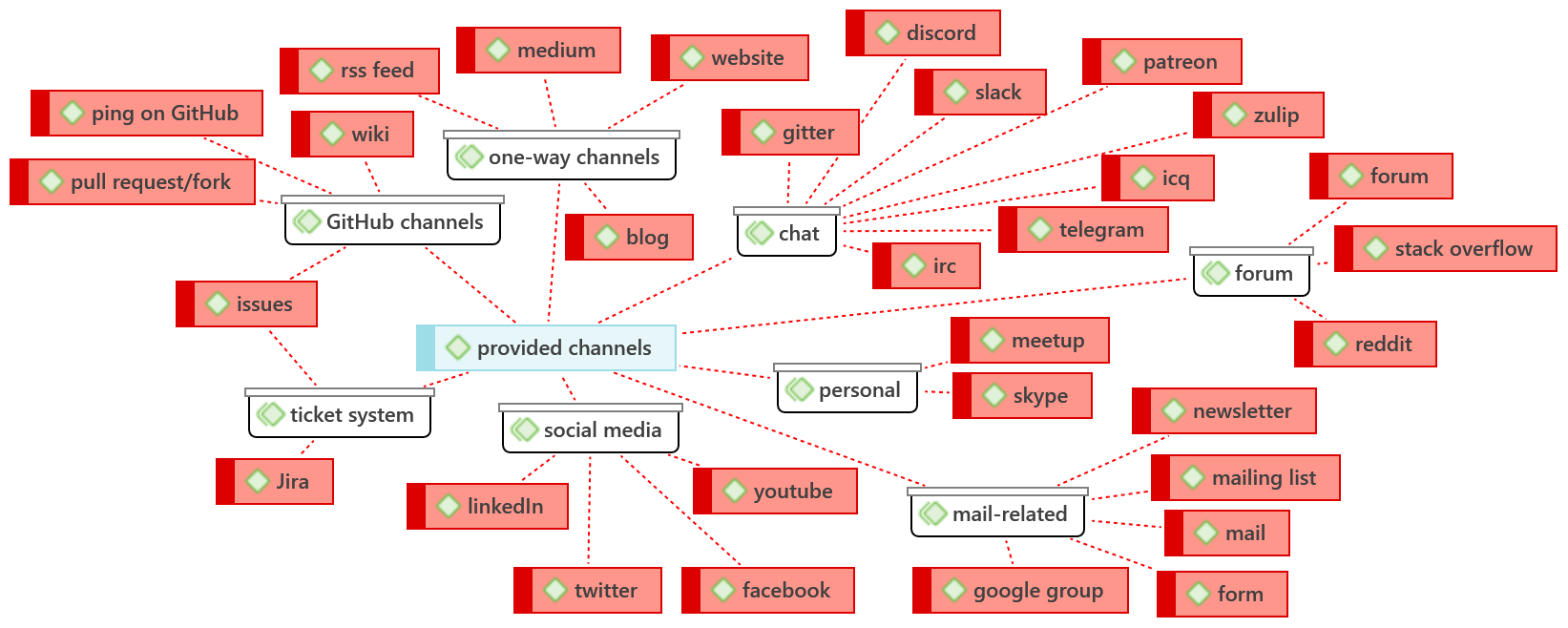}
\caption{Provided channels}
\label{fig:channels}
\end{figure*}

Figure~\ref{fig:channels} shows every channel that was provided in at least one of the analyzed projects. Most projects used the GitHub-given communication channels like \textit{issue}, \textit{pull request}, \textit{wiki} or \textit{pinging a maintainer} directly on GitHub. Additionally, some projects used another ticket system like \textit{Jira}. Some projects were active on social media channels like \textit{Twitter} or \textit{Facebook}. Other types of communication channels were mail-related, for example \textit{newsletters} or simple \textit{mail addresses}. In some projects, there were personal meetings, either in person (\textit{meetup}) or via \textit{skype}. Forums were used, for example \textit{stack overflow} or \textit{Reddit}. Last, many projects provided chat channels such as \textit{slack} or \textit{IRC}. 

\subsection{Channel Properties}

\begin{figure*}[ht]
\centering
\includegraphics[width=0.85\linewidth]{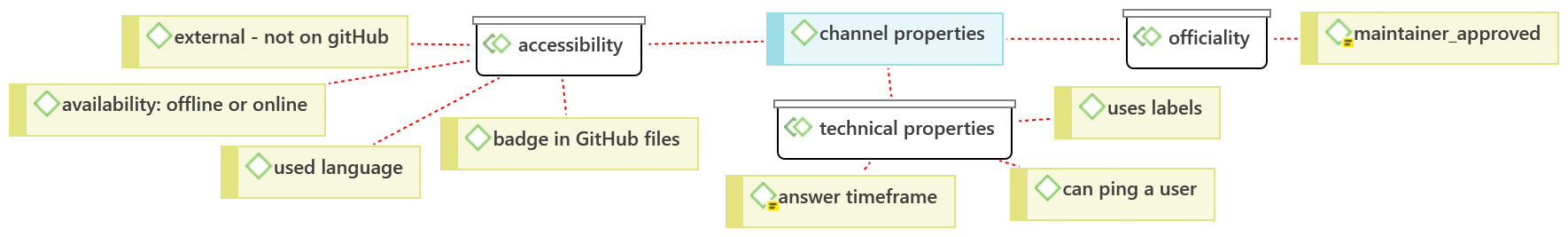}
\caption{Channel properties}
\label{fig:channel_properties}
\end{figure*}

Each channel has a set of properties, as can be seen in Figure~\ref{fig:channel_properties}. Some channels are \textit{maintainer approved}, meaning that some channels were officially created by the maintainers whereas other channels are unofficial.  

\begin{quote}
   \say{For chat, consider trying the \#general or \#beginners channels of the unofficial community Discord, the \#rust-usage channel of the official Rust Project Discord, or the \#general stream in Zulip.}~\footnote{\label{foot:serde}\url{https://github.com/serde-rs/serde/blob/master/README.md}}
\end{quote}

On the technical side, channels have different \textit{answer timeframes}, meaning they can be asynchronous or synchronous. 
\begin{quote}
     \say{The chat room is available for less formal and realtime discussion.}~\footnote{\label{foot:brian}\url{https://github.com/OpenDirective/brian/blob/master/README.md}}
\end{quote}

\begin{quote}
     \say{For asynchronous, consider the [rust] tag on StackOverflow, the /r/rust subreddit which has a pinned weekly easy questions post, or the Rust Discourse forum.}~\footref{foot:serde}
\end{quote}

Some channels \textit{use labels}, for example GitHub issues or Stack Overflow. 
\begin{quote}
     \say{Please ask usage and debugging questions on StackOverflow (use the "protractor" tag).}~\footnote{\url{https://github.com/juliemr/protractor/blob/master/README.md}}
\end{quote}

Other channels provide the possibility to \textit{ping a user}. 
\begin{quote}
    \say{Ping me @hectorerb in the IRC chatroom if you get stuck.}~\footnote{\label{foot:inventory}\url{https://github.com/flyve-mdm/ios-inventory-agent/blob/develop/README.md}}
\end{quote}

There are also different ways to access the channels. They can be \textit{off- or online}, they can be in different \textit{languages}, they can be accessed through a \textit{badge} on GitHub, and they can be \textit{external}, meaning not on GitHub. 

\subsection{Reasons for Channel Usage}
\label{sec:reasons}

\begin{figure*}[ht]
\centering
\includegraphics[width=0.85\linewidth]{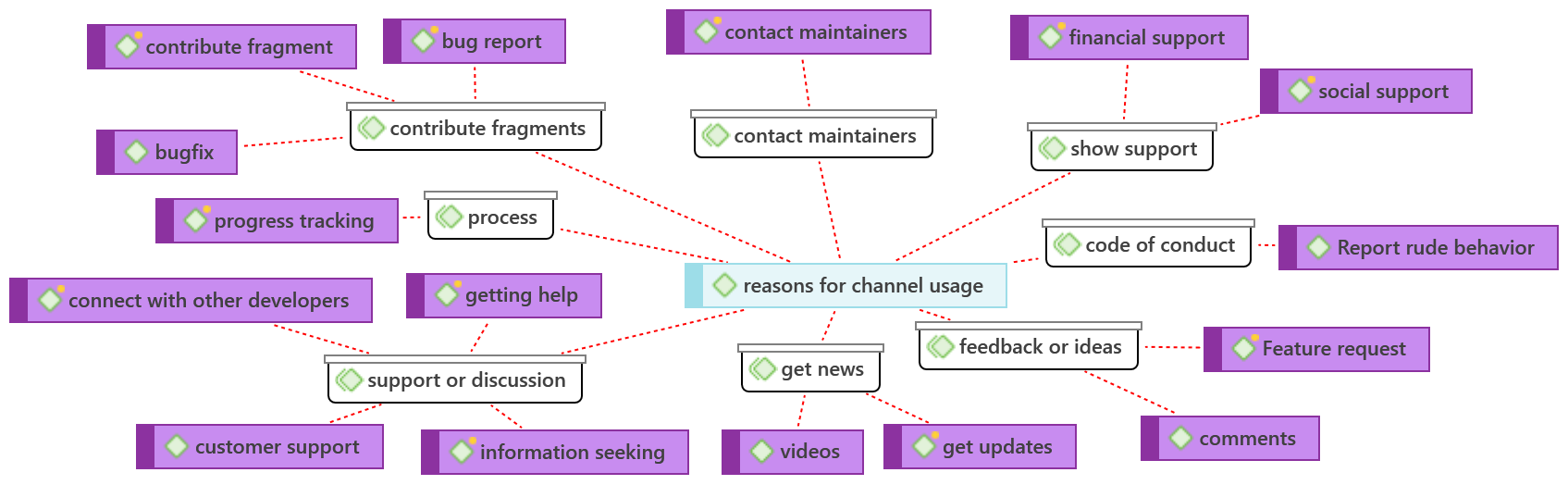}
\caption{Reasons for channel usage}
\label{fig:reasons}
\end{figure*}

Figure~\ref{fig:reasons} shows the reasons why one would use a communication channel. In open-source projects, there are channels used to \textit{contribute fragments}. Fragments can be improved documentation, new examples, translations, typo fixing, or code. Contributions can also be bug fixes or bug reports. 

\begin{quote}
     \say{Want to file a bug, contribute some code, or improve documentation?}~\footref{foot:inventory}
\end{quote}

\begin{quote}
    ``Here's some examples of things you might want to make a pull request for:
\begin{itemize}
    \item New language translations
    \item New features
    \item Bugfixes
    \item Inefficient blocks of code''~\footref{foot:inventory}
\end{itemize}
\end{quote}

Some channels can be used for \textit{progress tracking}. 
\begin{quote}
    \say{We use the Mailing list / group for general discussion plus issues on GitHub for detailed project discussion and progress tracking}~\footref{foot:brian}
\end{quote}

Other channels should be used if you need to \textit{get help}, \textit{seek information}, want to \textit{connect with other developers} or need \textit{customer support}.  
\begin{quote}
    \say{You can get help on our google group. Most support requests are answered very fast.}~\footnote{\url{https://github.com/chocoteam/choco-solver/blob/master/README.md}}
\end{quote}

\begin{quote}
     \say{Questions are free to be asked about the internals of the codebase and about the project.}~\footnote{\url{https://github.com/laravelio/laravel.io/blob/main/CONTRIBUTING.md}}
\end{quote}

\begin{quote}
   \say{To chat with the community and the developers we offer a Slack chat.}~\footnote{\url{https://github.com/arangodb/arangodb/blob/devel/README.md}}
\end{quote}

\begin{quote}
   \say{Use our official support channel.}~\footnote{\url{https://github.com/flyve-mdm/ios-inventory-agent/blob/develop/CONTRIBUTING.md}}
\end{quote}

One reason for a channel can be to \textit{get news}. The news can also be videos. 
\begin{quote}
    \say{Blog - A Week of A-Frame collects the latest news, upcoming events, and cool projects that other people are working on. The A-Frame Blog is also where you’ll find announcements and community case studies.}~\footnote{\url{https://aframe.io}}
\end{quote}

\begin{quote}
    \say{News and videos: [list of YouTube videos]}~\footnote{\url{https://developer.android.com/jetpack}}
\end{quote}

Another reason can be to \textit{give feedback or ideas} to the maintainers. This can be done via comments or as feature requests. 
\begin{quote}
    \say{For any comments, ideas, suggestions, issues, simply open an issue.}~\footnote{\url{https://github.com/layoutBox/PinLayout/blob/master/README.md}}
\end{quote}

As many channels follow a code of conduct, some channels can be used to\textit{ report rude behavior}. 
\begin{quote}
   \say{Instances of abusive, harassing, or otherwise unacceptable behavior may be reported by opening an issue or contacting one or more of the project maintainers.}~\footnote{\label{foot:dotfiles}\url{https://github.com/atomantic/dotfiles/blob/main/CODE_OF_CONDUCT.md}}
\end{quote}

\textit{Support} can be shown via social media. 
\begin{quote}
   \say{Star this repo to show support. Let me know you liked it on Twitter. Also, share the project.}~\footnote{\url{https://github.com/manrajgrover/halo/blob/master/README.md}}
\end{quote}

Finally, sometimes, one just wants to \textit{contact the maintainers}. 
\begin{quote}
  [translated]  \say{If somebody wants to join the team, please contact me using one of the following addresses.}~\footnote{\url{https://github.com/Kyohack/B2W2ger/README.md}}
\end{quote}

\subsection{How to Contribute}

\begin{figure*}[ht]
\centering
\includegraphics[width=\linewidth]{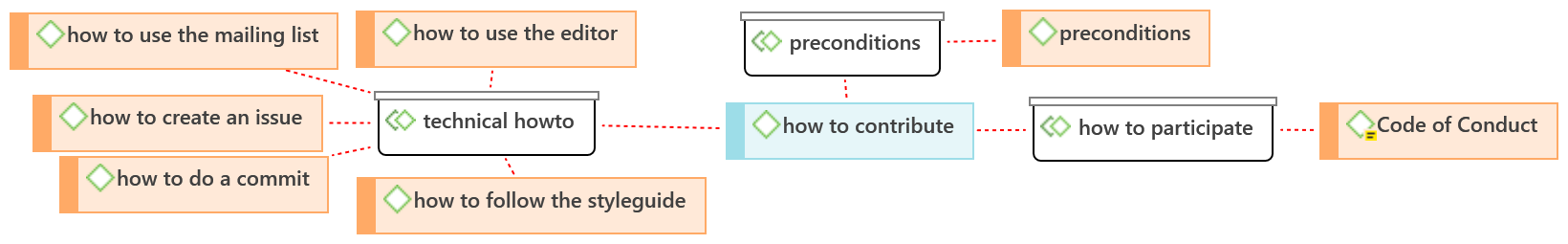}
\caption{How to contribute}
\label{fig:contribute}
\end{figure*}

As Figure~\ref{fig:contribute} shows, there are some \textit{preconditions} one has to fulfill before using some communication channels. 
\begin{quote}
 \say{Development of this plugin is done on GitHub. Pull requests welcome. Please see issues reported there before going to the plugin forum.}~\footnote{\url{https://github.com/GoogleChromeLabs/pwa-wp/blob/develop/README.md}}
\end{quote}

Users should follow a \textit{code of conduct} inside the channels. 
\begin{quote}
 \say{If any participant in this project has issues or takes exception with a contribution, they are obligated to provide constructive feedback and never resort to personal attacks, trolling, public or private harassment, insults, or other unprofessional conduct.}~\footref{foot:dotfiles}
\end{quote}

There are also some descriptions of how to use some channels, for example \textit{how to create an issue} or \textit{how to use a mailing list}.
\begin{quote}
 \say{Before filing a feature request, check the documentation to ensure it is not already provided.\\
Please provide the following information in an issue filed as a feature request:\\
What is the goal of the new feature?\\
If there is a current method of accomplishing this goal, describe the
problems or shortcomings of that method and how the proposed feature would improve the situation.}~\footref{foot:dotfiles}
\end{quote}
\begin{quote}
 \say{Please refer to the openSUSE Mailing Lists page to learn about our mailing list subscription and additional information.}~\footnote{\url{https://github.com/openSUSE/open-build-service/blob/master/README.md}}
\end{quote}

\subsection{Project Properties}
\begin{figure}[ht]
\centering
\includegraphics[width=\linewidth]{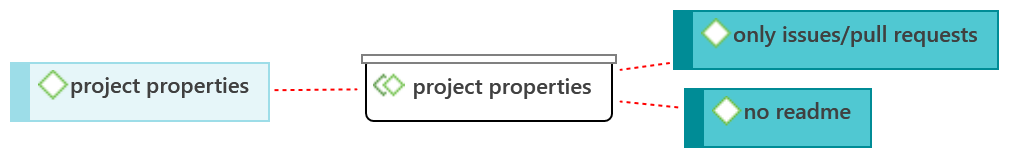}
\caption{Project properties}
\label{fig:project_properties}
\end{figure}

Some projects have the property of not officially offering communication channels. This can be seen in Figure~\ref{fig:project_properties}. These projects either have \textit{no README} at all or do not mention any communication channels in their documents. 

\section{Discussion} 
\label{sec:discussion}

In the following, we will  describe how our results answer our research questions. As Braun and Clarke describe, \say{thematic analysis has limited interpretative power beyond mere description if it is not used within an existing theoretical framework that anchors the analytic claims that are made}~\cite{Braun2006}. As described in the introduction, Storey et al. give six recommendations for maintainers that should be followed to prevent common challenges developers face. We want to use these recommendations as a framework around our results to better understand the consequences of how the channels are presented in GitHub projects.

\subsection{RQ 1: What Types of Communication Channels Do GitHub Projects Provide?}

All of our analyzed projects used communication channels. However, in 22 projects, the maintainers have not provided any additional communication channels than the ones provided automatically by GitHub. Furthermore, in three projects, there was no README or other Markdown file with any information about communication channels.
This leads to the conclusion that some maintainers either are not aware that additional communication channels might be useful, or they considered other channels and still think that issues and pull requests are enough channels for their project. 
In the other projects, however, the maintainers chose to provide additional channels to issues and pull requests. 

In our theme \textit{provided channels} (see Figure~\ref{fig:channels}), we can see that GitHub projects offer a wide variety of channels. In the examined projects, 32 unique channels were provided, with nine being the most channels additional to GitHub channels in a single project.

These channels can be clustered into nine sub-themes or types. Channels of different types serve different needs. One-way channels can be used for \say{broadcasting} information from one author to a broader audience, whereas chat channels are used for \say{real-time conversation between two or more people over the internet or another computer network}\footnote{\url{https://www.dictionary.com/browse/chat}}. A channel of one type usually cannot be replaced with a channel of another type without loosing some of its specific features. 

The channels of the same type offer similar functionality. For example, all mail-related channels use email as a main functionality, either in a kind of mailing-list or directly between two people. However, also channels of the same type cannot be easily interchanged most of the time. For instance, the affordances between channels of one type can differ a lot. Affordances are the conflicting strengths and weaknesses of a channel~\cite{storey}. To name one, mailing lists are public to the subscribers, whereas single emails are private. 

However, we have found three types of channel where the main functionality of the channels are very much alike, namely \textit{chat}, \textit{forum} and \textit{ticket system}. For these three types, also the affordances are similar for each channel of a type. Chat channels, for example, are all synchronous, ephemeral and textual. Most chats are also public or have public sub-channels that everybody can access. Due to similar functionality, maintainers use other criteria to chose between channels. Maintainers seem to provide channels that are used by many users, and some also give a choice between two or more channels to match the personal taste of the  developers.

\begin{quote}
   \say{Sign up to the A-Frame Slack to hang out and join the discussion. The Slack is pretty active with currently over 7000 people! [...] There’s also an \#aframe channel on the Supermedium Discord if you prefer that.}~\footnote{\url{https://aframe.io/community/}}
\end{quote}

 However, in many projects, there is no explanation why more than one channel of the same type is provided. 
 
 \begin{quote}
   \say{You can also chat with us via IRC in \#flyve-mdm on freenode or @flyvemdm on Telegram.}~\footref{foot:inventory}
\end{quote}

For maintainers, with such a large variety of channels, it might be hard to decide which channels they want to offer. Storey et al. recommend to \say{\textit{think lean when adopting new tools}}. Otherwise, many developers have issues with channel overload. Also, using many channels might lead to distractions and interruptions and also lead to fragmentation of information inside one project~\cite{storey}. In addition, developers might contribute to more than one GitHub project, which makes the number of channels that need to be used even higher. 

We have seen that different types of channels serve different needs. Therefore, in a large project with many needs, there might be many channels. However, there are channels which fulfill the same need. Especially for forums, chats and issue tracker, maintainers should reconsider, if two or more channels of the same type are really necessary. 

Nevertheless, maintainers should not be afraid to adopt new technologies. Storey et al. recommend to \say{\textit{stay abreast of the latest tools that may improve development productivity and team work}}. New tools might aggregate communication from different tools in one channel, which can help with channel overload~\cite{storey}. We can see that many newer communication channels are provided in GitHub projects, such as Slack or Discord. Moreover, we found many “classic” communication channels like mailing lists or forums. However, there are also projects using old channels like ICQ. 
Tantisuwankul et al. have shown that the used GitHub channels change and evolve over time~\cite{Tantisuwankul2019}. It looks like this is the same for other communication channels as well. Maintainers are aware of recent tools and adapt them in their projects. 

However, there are also maintainers keeping to old channels. In general, maintainers have to find the balance between keeping up to date with channels and not adopting too many channels. 

With such a great variety of possible channels, even of one single type, we need to know which channels developers and maintainers like to most/least and why. 

Independent of the type of channel, we can see that the channels named as most important in Storey's study are also provided in our analyzed GitHub projects. We have found projects suggesting face-to-face meetings, Q\&Q sites (e.g., Stack Overflow), (micro)blogs, and chats among others. However, some channels named as important were not mentioned at all in our analyzed projects. The developers in the survey named books or rich content as important communication channels, which were not mentioned at all in our examined projects. However, the missing channels are channels that are not between maintainers and developers or among developers. They are channels created between a third party (the book author, for instance) and therefore cannot be expected to be mentioned in a GitHub project.

\begin{framed}
GitHub projects provide 32 unique channels of nine different types, whereby channels of different types serve different needs. However, there are projects with more than one channel fulfilling the same need. Therefore, maintainers have to consider which channels to provide and if more than one channel fulfilling the same need is really necessary. 
\end{framed}

\subsection{RQ 2: How Are Communication Channels on GitHub Presented to Their Intended Audience?} 

In our analyzed projects, we can see that channels have different properties (see Figure~\ref{fig:channel_properties}). The properties are used in three different ways. The technical properties \textit{can ping a user} and \textit{uses labels} are actively used by the maintainers to ask developers to use a technical feature of a channel. The maintainers were aware of these properties, but we have not seen any hints that these features were a reason for choosing one channel over another channel without this feature. 

In contrast, the two properties \textit{answer timeframe} (synchronous vs asynchronous) and \textit{maintainer-approved} were used to distinguish between two channels. 

 \begin{quote}
   \say{For chat, consider trying the \#general or \#beginners channels of the unofficial community Discord [...]. For asynchronous, consider the [rust] tag on StackOverflow [...].}~\footref{foot:serde}
\end{quote}

According to Storey et al., developers need to recognize the tensions between conflicting affordances~\cite{storey}. Therefore, they recommend to \say{\textit{be aware of channel affordances and choose tools accordingly}}. The answer timeframe is such an affordance. 

However, we did not find any written evidence that maintainers chose a channel because of a specific affordance. Also, in all examined projects, we found only one affordance described in the documents, whereas Storey et al. describe five other channel affordances, for example, private versus public, that are not explicitly mentioned in any of the analyzed projects. We expect maintainers to know the difference between synchronous and asynchronous channels~\cite{Holmstrom2006, storey}. Therefore, it might be possible that maintainers were aware of the differences between synchronous and asynchronous without being aware of the concept of channel affordances in general. As future work, we want to know if maintainers are aware of channel affordances and how this influences their choice of provided channels. 

The last set of properties (\textit{accessibility}) was not explicitly mentioned by the maintainers, but influenced by them. Maintainers decide where they want to host a channel and in what language the communication should be. Maintainers are also responsible to update the documentation to remove offline channels or add badges for their channels.

This shows that maintainers do not only have to choose which channels they want to provide, depending on functionality and affordances. They also have to choose carefully what language a channel should be in and how and where it can be accessed. They have to maintain the channels and the respective documentation while they are used. During our analysis, we have seen some projects where maintainers failed to remove offline channels or where the channel language was different from English. A different language is not a problem per se, but maintainers have to be aware that this might exclude potential developers. For future work, we would like to see if maintainers are aware of these channel maintenance tasks and how they chose the basic settings like the language. 

In our theme \textit{reasons for channel usage}, we found that maintainers give reasons for channel usage (see Figure~\ref{fig:reasons}. In our examined projects, maintainers gave 16 unique reasons for channel usage. In 36 of the analyzed 90 projects, maintainers provided at least two channels additional to the GitHub channels. All of these maintainers, except two, gave reasons when to use which channel.

Maintainers seem to be aware that different channels can be used for different reasons and that developers need to be aware of these reasons. Some reasons are often connected to a certain set of channels throughout all examined projects. For example, \textit{contributing fragments} and \textit{feature requests} are clearly connected to \textit{issues} or \textit{pull requests}. Also, \textit{support or discussion} should often be done via all kinds of \textit{chat}, \textit{Stack Overflow} or \textit{issue}. \textit{Reporting rude behavior} should be done via e-mail. \textit{Social support} should be done using \textit{social media}. 

In most projects that provided more than one additional channel, we could see that the provided channels were of different types. In the couple of projects that offered more than one channel of the same type, especially of chat, forum, or ticket system (see previous section), the reasons given for the channels of the same type were at least in parts the same.  

Storey et al.'s recommend to \say{\textit{define a communication covenant with project members}}. This covenant should describe which channel should be used for which activity~\cite{storey}. 
The maintainers giving reasons for a channel is a first step towards a communication covenant. However, if the same reason is used for several channels, developers cannot know for sure which channel they should use. Maintainers need to be aware of this and make it explicit for which reason which channel should be used, without using the same reason for multiple channels. 

Our theme \textit{how to contribute} shows that some maintainers gave explanations on how to use some of the channels (see Figure~\ref{fig:contribute}). In most cases, these explanations were connected to issues and pull requests. For example, maintainers explained the GitHub workflow or gave examples of how a good issue should look like. Also, the preconditions were mostly about pull requests and issues, for example running tests before a pull request or checking for duplicates before submitting an issue. 
On the other hand, none of the \textit{chat}, \textit{forum} or \textit{social media} channels were connected with any explanation on how to use them. 

It is important to be literate with the tools you use~\cite{storey}. Therefore, Storey et al. recommend to \say{\textit{take the time to learn how to use the channels most effectively}}. 
Getting literate is something the maintainers can influence only slightly. However, giving explanations for channels might help developers to get literate with the given channels. If channel users are not literate with a channel, it can lead to frustration of the other users~\cite{storey}. From our results, we can see that maintainers give explanations for the GitHub channels. This is helpful for developers, especially for newcomers~\cite{Steinmacher_2014}. On the other hand, common channels like chat or forum are not explained at all. It is up to the developers to get literate with these channels. 

All in all, we have seen that most GitHub projects provide at least one additional communication channel to the GitHub channels. Maintainers give reasons for channel usage, but they might use the same reason for several channels. Maintainers give explanations for the GitHub channels to help developers get literate with them. However, there are no explanations for most other channels. 

There are projects where the maintainers write about channel affordances and give reasons in which case a channel should be used. But there are also numerous projects not giving such information.  We do not know what maintainers think who did not explicitly write about channel affordances or communication covenant. Perhaps, they chose the given channels because of their personal preferences or because they do not know other channels and not based on affordances or other properties at all. 
We also do not know how developers think about the presentation of communication channels in GitHub projects. Are they aware of channel affordances? What do they think about the given reasons for a channel? Are explanations or introductions missing?
Finally, except for some customer support channels, most channels are made for developers. How do users perceive the channels?

All in all, there is a clear connection from our results to the recommendations by Storey et al. although our coding was independent of it. It strengthens the themes identified in both studies.

\begin{framed}
Maintainers provide details about how and when to use a channel. However, there is often a lack of information on why maintainers chose to provide one channel instead of another and if they are aware of the differences or affordances of channels.
\end{framed}

\subsection{Other Observations}
During our analysis, we saw that most projects have a similar set of files where communication channels are mentioned, namely the markdown files README, CONTRIBUTING and CODE OF CONDUCT. 

There seems to be some common ground among maintainers about which files should be included in a GitHub project. Having a standardized set of files helps developers to find all necessary information for contributing to a project. It also helps to structure the information. 

\balance

\subsection{Implications}
Our results show that there are maintainers who are aware of channel properties and give reasons in which cases a channel should be used.  However, there are also projects without such information. Also, some projects provided numerous channels, which might be overwhelming. This leads to the implication that at least some project maintainers should check their channel setup to see if they can add additional information and if they can reduce the number of provided channels.

To do so, maintainers can use our results.  They can compare our results with their project to see what information could be included. Do they provide channels of the same type? Do they provide more than one channel for the same reasons? How many communication channels does their project provide? What affordances do these channels have? Are the documents about my channels up to date, or do they include offline channels? Is there a reason for channel usage that none of the provided channels can fulfill?

We do not know why maintainers chose the communication channels they provide in their projects. It would be interesting to explore the reasons and feeling of the maintainers towards the provided communication channels. If maintainers are unhappy with the current situation, guidelines for a good set of communication channels and how to introduce them might help. If maintainers are happy with their setup, it might help to raise awareness about how the developers feel about the setup and also look for a setup that fulfills the needs of both sides, developers and maintainers. In either way, maintainers have to chose wisely which channels they offer.

\subsection{Limitations}

First, as developers ourselves, we were not completely free of previous knowledge about how GitHub projects work and communicate. However, none of the authors has previously analyzed the written information about GitHub communication channels before or actively looked into the details about the channels. Therefore, we could still do the coding as open-minded as needed. 
Second, the list of large GitHub projects we used was last updated in 2019. Therefore, there might be some newer projects that were not part of that list. On the other hand, by analyzing projects that have existed for at least two years, we ensured that these projects had time to set up communication channels in case they required it.
Third, we have only analyzed what was written down in the GitHub documents. If a maintainer did not explicitly write about something, say, channel affordances, we no dot know if the concept was unknown or if the maintainer chose not to write about it. We want to further investigate this in future work.

\section{Conclusions and Future Work}

In our study, we have analyzed 90 GitHub projects to see how communication channels are presented in the written documents. We have used thematic analysis and found five themes, namely \textit{provided channels}, \textit{channel properties}, \textit{reasons for channel usage}, \textit{how to contribute} and \textit{project properties}. The most common channels were chats, mail-related channels, social media channels or GitHub channels. Channels can have several properties, for example, how they can be accessed or technical properties like the answer timeframe. There are many reasons for which a channel should be used. Channels can be used to contribute fragments, to get support or discuss things, or to give feedback or ideas among other reasons. To use a channel, some preconditions might have to be fulfilled, and some channels have explanations on how to use them. Although many projects do offer communication channels, some projects do not provide any communication channels or README file. 

Our analysis only focuses on what information is given by the maintainers in the written documentation. We have no information so far about what GitHub users think and feel about this topic. 

For future work, we want  to further analyze what developers and maintainers think about the given channels. Are they aware of channel affordances? Which channels do they like the least or the most? How do maintainers choose which channels they want to provide? 

We want to conclude interviews and/or questionnaires with GitHub users to answer the research questions proposed in the previous section. 
The results should then be combined with our current results to form a theory on communication in GitHub projects.

\subsection*{Data Availability}
We share raw data and scripts openly~\cite{verena_2022_5909440}. 

\bibliographystyle{ACM-Reference-Format}
\bibliography{ref}

\end{document}